\documentclass[fleqn,usenatbib]{mnras}

\usepackage{newtxtext,newtxmath}

\usepackage[T1]{fontenc}
\usepackage{ae,aecompl}
\usepackage{lscape}

\usepackage{graphicx}	
\usepackage{amsmath}	
\usepackage{amssymb}	

\newcommand{\NGTS}{NGTS}
\newcommand{\TESS}{{\it TESS}}

\newcommand{\kms}{km\,s$^{-1}$}
\newcommand{\ms}{m\,s$^{-1}$}
\newcommand{\masy}{mas\,y$^{-1}$}
\newcommand{\mpl}{\mbox{M$_{P}$}}
\newcommand{\rpl}{\mbox{R$_{P}$}}
\newcommand{\rhopl}{\mbox{$\rho_{P}$}}
\newcommand{\mstar}{\mbox{M$_{*}$}}
\newcommand{\rstar}{\mbox{R$_{*}$}}
\newcommand{\rhostar}{\mbox{$\rho_{*}$}}
\newcommand{\mjup}{\mbox{M$_{J}$}}
\newcommand{\rjup}{\mbox{R$_{J}$}}
\newcommand{\rhojup}{\mbox{$\rho_{J}$}}
\newcommand{\mearth}{\mbox{M$_{\oplus}$}}
\newcommand{\rearth}{\mbox{R$_{\oplus}$}}
\newcommand{\msun}{\mbox{M$_{\odot}$}}
\newcommand{\rsun}{\mbox{R$_{\odot}$}}
\newcommand{\gccc}{g\,cm$^{-3}$}
\newcommand{\ergscm}{erg\,s$^{-1}$cm$^{-2}$}
\newcommand{\teff}{$T_{\rm eff}$}
\newcommand{\feh}{\mbox{[Fe/H]}}
\newcommand{\vsini}{\mbox{$v\sin i$}}
\newcommand{\logg}{$\log g$}
\newcommand{\tc}{$T_C$}

\newcommand{\lnL}{\mbox{$\ln\mathcal{L}$}}
\newcommand{\Nstar}{NGTS-12}
\newcommand{\Nticid}{TIC-157230659}
\newcommand{\Ntwomass}{2MASS J11450000-3538261}
\newcommand{\NGAIAid}{3464620139089832960}

\newcommand{\Nkamp}{\mbox{$21.3\pm2.1$}}
\newcommand{\Nlogk}{\mbox{$3.06 \pm 0.10$}}
\newcommand{\Ngamma}{\mbox{$-12595.6 \pm1.4$}}
\newcommand{\Noffsetferos}{\mbox{$-3.0\pm7.1$}} 
\newcommand{\NRA}{\mbox{$11^{\rmn{h}} 44^{\rmn{m}} 59\fs986$}} 
\newcommand{\NDec}{\mbox{$-35\degr 38\arcmin 26\farcs 0267$}} 
\newcommand{\Nplx}{\mbox{$2.212 \pm 0.038$}}
\newcommand{\NpropRA}{\mbox{$-18.4141\pm0.0441 $}} 
\newcommand{\NpropDec}{\mbox{$7.3178\pm0.0320$}} 
\newcommand{\Nstarmass}{\mbox{$1.021\,^{+0.056}_{-0.049}$}} 
\newcommand{\Nstarradius}{\mbox{$1.589 \pm 0.040$}} 
\newcommand{\Nstardensity}{\mbox{$0.359\pm0.039$}} 
\newcommand{\Nstardensityglobal}{\mbox{$0.362\,^{+0.022}_{-0.032}$}} 
\newcommand{\NLDoneNG}{\mbox{$0.30\,^{+0.22}_{-0.19}$}} 
\newcommand{\NLDtwoNG}{\mbox{$0.35\,^{+0.31}_{-0.37}$}}
\newcommand{\NLDoneTS}{\mbox{$0.51\,^{+0.25}_{-0.27}$}}
\newcommand{\NLDtwoTS}{\mbox{$0.06\,^{+0.41}_{-0.31}$}}
\newcommand{\Nteff}{\mbox{$5690 \pm 130$}} 
\newcommand{\Nmetal}{\mbox{$-0.03\pm0.11$}} 
\newcommand{\Ndist}{\mbox{$452.0\pm7.6$}} 
\newcommand{\Nlogg}{\mbox{$4.045 \pm 0.038$}} 

\newcommand{\Nvsini}{$< 0.5$}
\newcommand{\Nage}{\mbox{$9.4\pm1.5$}} 

\newcommand{\NVmag}{$12.382 \pm 0.046$}
\newcommand{\NBmag}{$12.974 \pm 0.051$}

\newcommand{\NGAIAmag}{$12.1279 \pm 0.0002$}
\newcommand{\Nbpmag}{$12.4947 \pm 0.0008$}
\newcommand{\Nrpmag}{$11.6071 \pm 0.0012$}
\newcommand{\NNmag}{$11.83 \pm 0.01$}
\newcommand{\NJmag}{$10.992 \pm 0.026$}
\newcommand{\NHmag}{$10.692 \pm 0.024$}
\newcommand{\NKmag}{$10.588 \pm 0.021$}
\newcommand{\NWmag}{$10.555 \pm 0.023$} 
\newcommand{\NWWmag}{$10.597 \pm 0.020$}
\newcommand{\NTmag}{$11.6634 \pm 0.006$}
\newcommand{\NTmagshort}{$11.6$}
\newcommand{\NVmagshort}{$12.38$}

\newcommand{\Nplanet}{NGTS-12b}
\newcommand{\Nperiod}{\mbox{$7.532806\pm0.000048$}}
\newcommand{\Nperiodshort}{\mbox{$7.53$}}
\newcommand{\Nduration}{\mbox{$5.85 \pm 0.073$}}
\newcommand{\Ndepth}{\mbox{$0.55$}}

\newcommand{\Ntc}{\mbox{$2458572.8549\pm0.0017$}}
\newcommand{\Necc}{\mbox{$0.01^{+0.03}_{-0.01}$}}%
\newcommand{\Nmass}{\mbox{$0.208\pm0.022$}}%
\newcommand{\Nradius}{\mbox{$1.048\pm0.032$}}%
\newcommand{\Ndensitycgs}{\mbox{$0.223\pm0.029$}} 
\newcommand{\Ngravp}{\mbox{$2.670\,^{+0.049}_{-0.056}$}}
\newcommand{\Nflux}{\mbox{($5.66\pm0.62) \times 10^{8}$}}%
\newcommand{\Nrratio}{\mbox{$0.0678\pm0.0012$}}
\newcommand{\Nau}{\mbox{$0.0757\pm0.0014$}}%
\newcommand{\Naoverr}{\mbox{$10.28\,^{+0.21}_{-0.31}$}}%

\newcommand{\Nimpact}{\mbox{$0.20\pm0.13$}}%
\newcommand{\Ninc}{\mbox{$88.90\pm0.76$}}
\newcommand{\Ntaucirc}{\mbox{$3.77\pm0.90$}} 
\newcommand{\NdilutionMC}{\mbox{$8.5\pm4.7$}} 
\newcommand{\NdilutionCalc}{\mbox{$6.09\pm0.15$}}
\newcommand{\NmeanNG}{\mbox{$1.000168\pm0.000019$}}
\newcommand{\NsigNG}{\mbox{$0.001996\pm0.000018$}}
\newcommand{\NmeanTS}{\mbox{$0.999985\pm0.000026$}}
\newcommand{\NTeq}{\mbox{$1257\pm34$}}
\newcommand{\NHkm}{\mbox{$1020\,^{+150}_{-130}$}} 

\title[\Nplanet]{\Nplanet: A sub-Saturn mass transiting exoplanet in a \Nperiodshort\,day orbit}

\author[E. Bryant et al.]{
\parbox{\textwidth}{
Edward M.~Bryant,$^{1,2}$\thanks{E-mail: \href{edward.bryant@warwick.ac.uk}{edward.bryant@warwick.ac.uk}}
Daniel~Bayliss,$^{1,2}$
Louise D. Nielsen,$^{3}$
Dimitri Veras$^{1,2}$\thanks{STFC Ernest Rutherford Fellow}
Jack S.~Acton,$^{4}$
David~R.~Anderson,$^{1,2}$
David J. Armstrong,$^{1,2}$\footnotemark[2]
Fran\c{c}ois Bouchy,$^{3}$
Joshua~T.~Briegal,$^{8}$
Matthew R. Burleigh,$^{4}$
Juan~Cabrera,$^{5}$
Sarah L. Casewell,$^{4}$\footnotemark[2]
Alexander Chaushev,$^{6}$
Benjamin F. Cooke,$^{1,2}$
Szil\'ard~Csizmadia,$^{5}$
Philipp~Eigm\"uller,$^{5}$
Anders~Erikson,$^{5}$
Samuel Gill,$^{1,2}$
Edward~Gillen,$^{7,8}$\thanks{Winton Fellow}
Michael R.~Goad,$^{4}$
Nolan Grieves,$^{3}$
Maximilian~N.~G{\"u}nther,$^{9}$\thanks{Juan Carlos Torres Fellow}
Beth Henderson,$^{4}$
Aleisha Hogan,$^{4}$
 James~S.~Jenkins,$^{10,11}$
Monika Lendl,$^{3}$
James~McCormac,$^{1,2}$
Maximiliano~Moyano,$^{12}$
Didier~Queloz,$^{8}$
Heike~Rauer,$^{5,6}$
Liam Raynard,$^{4}$
Alexis~M.~S.~Smith,$^{5}$
Rosanna~H.~Tilbrook,$^{4}$
St\'{e}phane~Udry,$^{3}$
Jose I. Vines$^{10}$
Christopher~A.~Watson,$^{13}$
Richard~G.~West,$^{1,2}$
Peter~J.~Wheatley,$^{1,2}$
}
\\
$^{1}$Dept.\ of Physics, University of Warwick, Gibbet Hill Road, Coventry CV4 7AL, UK\\
$^{2}$Centre for Exoplanets and Habitability, University of Warwick, Gibbet Hill Road, Coventry CV4 7AL, UK\\
$^{3}$Observatoire de Gen{\`e}ve, Universit{\'e} de Gen{\`e}ve, Chemin des Maillettes 51, 1290 Sauverny, Switzerland\\
$^{4}$School of Physics and Astronomy, University of Leicester, Leicester, LE1 7RH, UK\\
$^{5}$Institute of Planetary Research, German Aerospace Center, Rutherfordstrasse 2, 12489 Berlin, Germany\\
$^{6}$Center for Astronomy and Astrophysics, TU Berlin, Hardenbergstr. 36, D-10623 Berlin, Germany\\
$^{7}$Astronomy Unit, Queen Mary University of London, Mile End Road, London E1 4NS, UK\\
$^{8}$Astrophysics Group, Cavendish Laboratory, J.J. Thomson Avenue, Cambridge CB3 0HE, UK\\
$^{9}$Department of Physics, and Kavli Institute for Astrophysics and Space Research, Massachusetts Institute of Technology, Cambridge, MA 02139, USA\\
$^{10}$Departamento de Astronomia, Universidad de Chile, Casilla 36-D, Santiago, Chile\\
$^{11}$ Centro de Astrof\'isica y Tecnolog\'ias Afines (CATA), Casilla 36-D, Santiago, Chile.\\
$^{12}$Instituto de Astronom\'ia, Universidad Cat\'{o}lica del Norte, Casa Central, Angamos 0610, 1270709, Antofagasta, Chile\\
$^{13}$Astrophysics Research Centre, School of Mathematics and Physics, Queen's University Belfast, BT7 1NN Belfast, UK\\
}

\date{Accepted 2020 September 21. Received 2020 September 18; in original form 2020 August 17
}

\pubyear{2020}

\begin{document}
\label{firstpage}
\pagerange{\pageref{firstpage}--\pageref{lastpage}}
\maketitle

\begin{abstract}
We report the discovery of the transiting exoplanet \Nplanet\ by the Next Generation Transit Survey (\NGTS). The host star, \Nstar, is a V=\NVmagshort\,mag star with an effective temperature of T$_{\rm eff}$=\Nteff\,K. \Nplanet\ orbits with a period of $P=\Nperiodshort$\,d, making it the longest period planet discovered to date by the main \NGTS\ survey. We verify the \NGTS\ transit signal with data extracted from the TESS full-frame images, and combining the photometry with radial velocity measurements from HARPS and FEROS we determine \Nplanet\ to have a mass of \Nmass\,\mjup\ and a radius of \Nradius\,\rjup. \Nplanet\ sits on the edge of the Neptunian desert when we take the stellar properties into account, highlighting the importance of considering both the planet and star when studying the desert. The long period of \Nplanet\ combined with its low density of just \Ndensitycgs\,\gccc\ make it an attractive target for atmospheric characterization through transmission spectroscopy with a Transmission Spectroscopy Metric of 89.4.  
\end{abstract}




\section{Introduction}
\label{sec:intro}
Detecting exoplanets via transits has proved to be a very successful path for discovering other worlds.  The process began with discoveries from the deep OGLE \citep{udalski1992ogle} survey \citep[eg.][]{konacki2003ogle1, bouchy2004ogle2}.  Soon discoveries were being made via purpose built wide-field survey cameras such as WASP \citep{pollacco2006WASP}, HATNet \citep{bakos2004hatnet}, and KELT \citep{pepper2007kelt}.  Following these successes, space-based missions, primarily CoRoT \citep{baglin2006corot} and Kepler \citep{borucki2010kepler}, delivered high precision photometry which allowed for the discovery of smaller radius planets.  Currently the TESS mission \citep{ricker2014} is combining the advantages of wide field cameras and space-based photometric precision to uncover transiting exoplanets orbiting bright stars over almost the entire sky.

The Next Generation Transit Survey \citep[\NGTS;][]{wheatley18ngts} is a ground-based exoplanet hunting facility which is situated at ESO's Paranal Observatory in Chile. It consists of twelve fully robotic telescopes each with a 20\,cm photometric aperture and a wide field-of-view of 8\,deg$^2$.  By combining the excellent observing conditions at Paranal Observatory with back-illuminated CCD cameras and sub-pixel level autoguiding \citep{mccormac2013donuts} on ultra-stable mounts, \NGTS\ can achieve a higher photometric precision than previous ground-based facilities.  This was demonstrated with the discovery of NGTS-4b \citep{west2019ngts4}, a Neptune-sized exoplanet with a transit depth of just 0.13\%.  \NGTS\ provides a useful complement to the \TESS\ survey, with a higher spatial resolution, a more flexible observing schedule, and better photometric precision for stars with T$>$13.

The majority of discoveries from ground-based transit surveys to date are exoplanets with periods shorter than 5 days.  This bias towards short period planets is a combination of the geometric probability of transit and the difficulty of conducting long duration photometric campaigns.  \NGTS\ is able to mitigate this difficulty in two ways.  Firstly, we can conduct longer monitoring campaigns due to the excellent site conditions at Paranal Observatory.  Secondly, with higher photometric precision, the signal-to-noise of each individual transit is higher, and so fewer transits are required to build up the same total signal-to-noise.

In this paper we report the discovery of \Nplanet; a \Nmass\,\mjup, \Nradius\,\rjup\ transiting exoplanet with a period of \Nperiodshort\,days. \Nplanet\ is the longest period planet discovered to date from the main \NGTS\ survey.  In Section~\ref{sec:obs} we detail the photometric and spectroscopic observations obtained of \Nstar.  In Section~\ref{sec:analysis} we perform a global analysis of the data to determine the stellar and planetary properties of the system.  We discuss our results in Section~\ref{sec:discussion} and finally present our conclusions in Section~\ref{sec:conc}.


\section{Observations}
\label{sec:obs}
\Nstar\ is a T=\NTmagshort\ star in the southern hemisphere (R.A.=$\NRA$, Dec.=$\NDec$) which was monitored as part of the \NGTS\ survey.  In this section we set out the photometric and spectroscopic observations of \Nstar\ that led to the discovery of the exoplanet \Nplanet. 

\subsection{\NGTS{} Photometry}
\label{sub:ngtsphot}
\begin{figure*}
	\includegraphics[width=\textwidth]{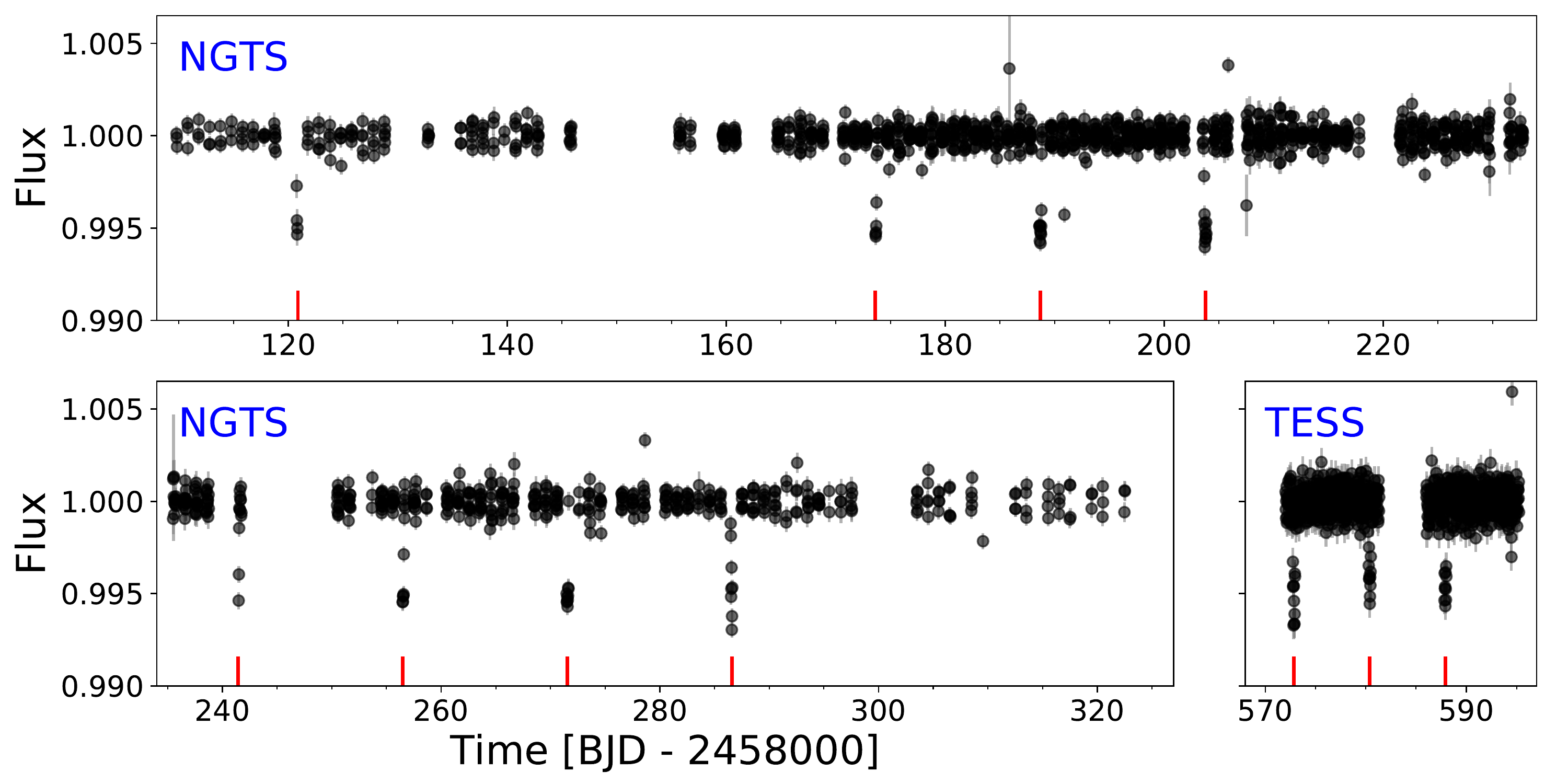}
    \caption{Full \NGTS\ and TESS light curves for \Nstar. The red vertical lines give the positions of the observed transits of \Nplanet. A total of 8 transit events are recorded in the \NGTS\ data, and 3 in the TESS data.  The \NGTS\ data has been binned to 30\,minutes and flattened in a similar manner to the TESS data for visual comparison with the 30\,minute cadence TESS data.}
    \label{fig:phot1}
\end{figure*}

\Nstar\ was observed using a single \NGTS\ telescope between 2017 December 10 and 2018 August 4.  During this period a total of 250,503 images were taken with an exposure time of 10\,seconds.  Image reduction and photometry was performed using the CASUTools\footnote{\url{http://casu.ast.cam.ac.uk/surveys-projects/software-release}} photometry package as detailed in \citet{wheatley18ngts}. The detrending of the light curve was performed using an implementation of the SysRem algorithm \citep{Tamuz2005} to remove signals which are exhibited by multiple stars across the field. The full \NGTS\ survey light curve is shown in Fig.~\ref{fig:phot1}, and the photometry is provided in Table~\ref{tab:phot}.

The \Nstar\ light curve was searched for periodic, transit-like signals using \texttt{ORION}, a custom implementation of the Box-fitting Least Squares algorithm \citep[BLS;][]{Kovacs2002}. A strong peak in the BLS periodogram was detected at \Nperiodshort\,days. Folding the \NGTS\ photometry on this period revealed a clear transit signal with a depth of \Ndepth\% and a duration of \Nduration\,hours.

In GAIA DR2 \citep{GAIA_DR2} there are no other stars within the photometric radius (15\,\arcsec) of \Nstar.\@ The parallax and photometric measurements from GAIA DR2 (listed in Table~\ref{tab:stellar}) also rule out the possibility that \Nstar\ is a giant star.  The lightcurve for \Nstar\ shows no evidence of a secondary eclipse, odd-even differences in the transit depths, or out-of-transit variability.  A convolutional neural network implemented for NGTS data \citep{chaushev19cnn} reported a probability of $> 0.99$ for the transit signal being planetary in nature.  We therefore concluded that \Nplanet\ was a strong transiting exoplanet candidate, and initiated spectroscopic follow-up detailed in Sections \ref{sub:speccoralie} to \ref{sub:spectferos}.

\begin{table}
	\centering
	\caption{\NGTS\ and TESS photometry for \Nstar.  The full table is available in a machine-readable format from the online journal.  A portion is shown here for guidance.}
	\label{tab:phot}
	\begin{tabular}{ccccc}
	Time	&	Flux        	&Flux    &    &Exp. Time\\
    (BJD-2450000)	&	(normalised)	&error  &Instrument  &(s)\\
	\hline
    8097.82554  &  1.0092  &  0.0056  &  \NGTS  & 10 \\
    8097.82569  &  0.9879  &  0.0055  &  \NGTS  & 10 \\
    8097.82584  &  0.9899  &  0.0055  &  \NGTS  & 10 \\
    8097.82599  &  0.9969  &  0.0055  &  \NGTS  & 10 \\
    8097.82614  &  0.9918  &  0.0055  &  \NGTS  & 10 \\
    8097.82628  &  1.0120  &  0.0055  &  \NGTS  & 10 \\
    8097.82643  &  0.9885  &  0.0055  &  \NGTS  & 10 \\
    8097.82658  &  1.0046  &  0.0055  &  \NGTS  & 10 \\
    8097.82673  &  1.0085  &  0.0055  &  \NGTS  & 10 \\
    8097.82689  &  0.9945  &  0.0055  &  \NGTS  & 10 \\
    ...         &  ...     &  ...     & ...    & ... \\
    8572.01562  &  1.0005  &  0.0008  &  TESS  & 1800 \\ 
    8572.03644  &  0.9996  &  0.0008  &  TESS  & 1800 \\ 
    8572.05731  &  0.9991  &  0.0008  &  TESS  & 1800 \\ 
    8572.07812  &  1.0000  &  0.0008  &  TESS  & 1800 \\ 
    8572.09894  &  1.0004  &  0.0008  &  TESS  & 1800 \\ 
    8572.11981  &  1.0008  &  0.0008  &  TESS  & 1800 \\ 
    8572.14062  &  0.9988  &  0.0008  &  TESS  & 1800 \\ 
    8572.16144  &  0.9992  &  0.0008  &  TESS  & 1800 \\ 
    8572.18231  &  1.0001  &  0.0008  &  TESS  & 1800 \\ 
    8572.20312  &  1.0002  &  0.0008  &  TESS  & 1800 \\ 
    ...         &  ...     &  ...     &  ...   & ... \\

	\hline
	\end{tabular}
\end{table}


\subsection{TESS Photometry}
\label{sub:tessphot}
\begin{figure}
    \centering
    \includegraphics[width=\columnwidth]{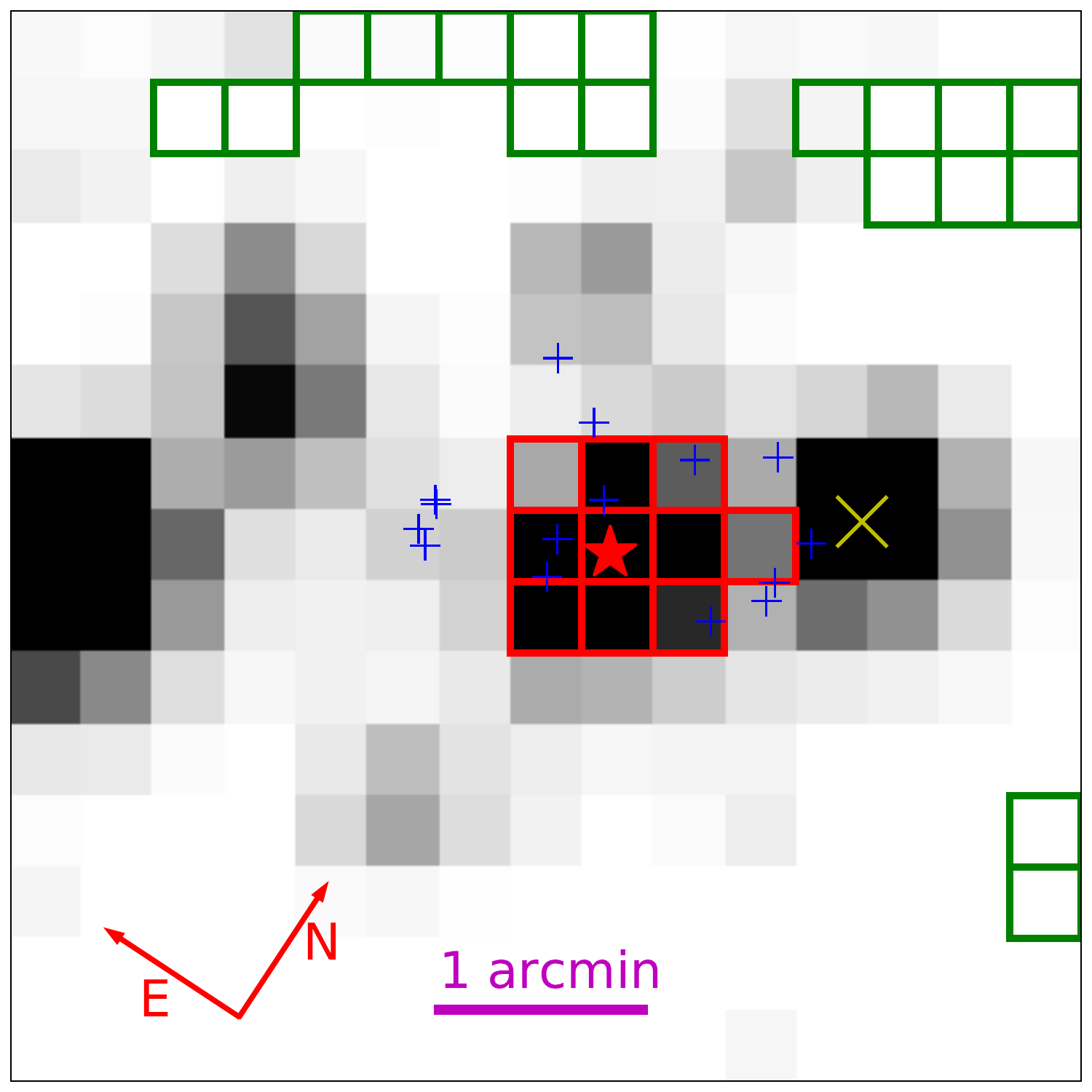}
    \caption{TESS Full-Frame Image cutout ($15\times15$ pixels) in the region of \Nstar\ from which the TESS light curve was generated. The red boxes indicate the target pixel aperture mask used to extract the photometry and the green boxes denote the pixels used to determine the sky background. The red star indicates the position of \Nstar, the yellow cross gives the location of TIC157230670.  The blue pluses denote the positions of faint neighbouring stars (T$>$15.5) detected in GAIA DR2 within 1\,\arcmin\ of \Nstar, used for dilution estimation (Section~\ref{sub:dilution}).}
    \label{fig:tessap}
\end{figure}
One year after the \NGTS\ monitoring set out in Section~\ref{sub:ngtsphot}, \Nstar\ was also observed during Sector 10 (Camera 2; CCD 4) of the TESS mission \citep{ricker2014} between 2019 March 28 and 2019 April 21.  \Nstar\ was not selected for 2-minute TESS photometry.  We therefore created a lightcurve via aperture photometry of the TESS 30-minute cadence full-frame images (FFIs) following the method set out in \citet{gill20}.  Postage stamps of 15x15 pixels were downloaded for all Sector 10 TESS FFIs.  We determined a threshold for target pixels and for background pixels based on an iterative sigma-clipping to determine the median and standard deviation of the background pixels.  To exclude the neighbouring $T=12.0$ star (TIC157230670, 76\arcsec\ to the north-west, see Fig.~\ref{fig:tessap}) as much as possible, we only include pixels in our final aperture for which neighbouring pixels closer to the centre of our target show a higher illumination. Otherwise we assume the pixel is influenced in majority by another source.  For the final pixel mask of \Nstar, 10 pixels over a threshold of 168 electrons/second were selected (see Fig.~\ref{fig:tessap}).  To remove the systematic trends from the TESS light curve, we mask out the transits, and then fit a spline with nodes spaced by 10\,hours, interpolating across the positions of the transits.

The TESS photometry for \Nstar\ is shown in Fig.~\ref{fig:phot1}, and clearly shows three transit events with depths, durations, and epochs consistent with the \NGTS\ data.  Again we see no evidence of secondary eclipses, odd-even differences in the transit depths, or out-of-transit variability. The TESS photometry is provided in Table~\ref{tab:phot}.

\subsection{CORALIE Spectroscopy}\label{sub:speccoralie}
Two reconnaissance spectra were obtained for \Nstar\ using the CORALIE spectrograph \citep{CORALIE} mounted on the Swiss-1.2\,m Leonard Euler telescope, located at the ESO La Silla Observatory, Chile. An exposure time of 2700\,s was used for both observations. The spectra were reduced using the standard CORALIE data reduction pipeline. The radial velocities (RVs) were extracted via cross-correlation with a G2 binary mask and are given in Table~\ref{tab:rvs}.
The RV measurements obtained from these spectra were useful in ruling out stellar mass companions. However, the mean uncertainty of the two RVs from these spectra is 34\,ms$^{-1}$, which is roughly 1.5 times the RV semi-amplitude for \Nplanet. As such, the CORALIE measurements do not have the precision to help constrain the planetary parameters, and so we do not include them in the global modelling detailed in Section~\ref{sub:global}.

\subsection{HARPS Spectroscopy}
\label{sub:spectharps}
We obtained multi-epoch spectroscopy for \Nstar\ with the HARPS spectrograph \citep{Mayor2003} between the dates of 2019 December 28 and 2020 February 11 under programme ID 0104.C-0588 (PI Bouchy). The HARPS spectrograph is mounted on the ESO 3.6\,m telescope at the La Silla Observatory, Chile. The spectra were reduced using the standard HARPS data reduction pipeline and RVs were extracted via cross correlation with a G2 binary mask. The full RV time series of eight measurements is given in Table~\ref{tab:rvs}. For all the HARPS observations an exposure time of 1800\,s was used. The typical RV uncertainty is 4 m/s.  We plot the phase-folded RV measurements in Fig.~\ref{fig:RV}, which clearly shows an RV variation in-phase with the photometric period determined in Section \ref{sub:ngtsphot}.  The RV semi-amplitude of $K=20$\,\ms is consistent with a transiting planet orbiting \Nstar.

We also extract the bisector spans of the HARPS cross correlation functions. We investigate the correlation between the RV measurements and the bisector spans and find a Pearson-R correlation coefficient of -0.21, indicating no strong correlation (see Fig.~\ref{fig:RV}). 

\subsection{FEROS Spectroscopy}\label{sub:spectferos}
We obtained two further spectroscopic observations of \Nstar\ with the FEROS spectrograph \citep{kaufer1999feros} on UT 2020 January 1 and 3, using an exposure time of 1200\,s. The FEROS spectrograph is mounted on the MPG/ESO-2.2\,m telescope at the La Silla Observatory, Chile. We reduced these spectra using the CERES pipeline \citep{CERES}. CERES also calculates RVs by cross-correlating the reduced spectra with a G2 binary mask and then fitting a double Gaussian model to the cross-correlation function in order to account for possible moonlight contamination. We note that the G2 mask used by the CERES pipeline is the same mask used to extract the CORALIE and HARPS RVs. The resulting RVs are listed in Table~\ref{tab:rvs} and shown in Fig.~\ref{fig:RV}.  

\begin{table}
	\centering
	\caption{Radial Velocities for \Nstar}
	\label{tab:rvs}
	\begin{tabular}{ccccc} 
BJD			&	RV		&RV err & Instrument & Exp. Time\\
	& (\kms)& (\kms) & & (s)\\
		\hline
2458826.809493&  -12.61315&  0.03399& CORALIE& 2700\\
2458830.809847&  -12.65738&  0.03343& CORALIE& 2700\\
2458845.783261&  -12.61483&  0.00835& HARPS& 1800\\
2458846.773136&  -12.61613&  0.00370& HARPS& 1800\\
2458847.794460&  -12.59168&  0.00295& HARPS& 1800\\
2458849.81984&   -12.5808&   0.0103&  \:FEROS\:& 1200\\
2458851.80857&   -12.6002&   0.0096&  \:FEROS\:& 1200\\
2458869.833151&  -12.60787&  0.00410& HARPS& 1800\\
2458871.854470&  -12.58030&  0.00357& HARPS& 1800\\
2458887.816510&  -12.57215&  0.00409& HARPS& 1800\\
2458889.835373&  -12.60464&  0.00453& HARPS& 1800\\
2458890.765724&  -12.61451&  0.00600& HARPS& 1800\\
		\hline
	\end{tabular}
\end{table}


\section{Analysis}
\label{sec:analysis}
Using the observations presented in Section~\ref{sec:obs}, along with stellar properties from a variety of astronomical surveys, we modelled the \Nplanet\ system.  We set out that analysis in this section.  


\subsection{Stellar Properties}
\label{sub:stellar}
We determined stellar atmospheric parameters of \Nstar\ using isochrone fitting with the \texttt{isochrones} Python module \citep{morton15isochrones}.  We obtained priors for \teff, \feh, \logg\ by spectral matching from the 1D stacked HARPS spectra (see Section~\ref{sub:spectharps}), using a custom wavelet analysis algorithm \citep{gill18wavelet} as well as \texttt{SpecMatch-emp} \citep{specmatchemp}. We took priors for these parameters that encompassed both spectral matching methods, and the resulting Gaussian priors were \teff~=~$5649\pm153$\,K, \feh~=~$-0.045\pm0.15$, and \logg~=~$3.95\pm0.5$.  For additional priors, we used the parallax, $B_P$ magnitude, and $R_P$ magnitude from GAIA DR2 \citep{GAIA_DR2} as listed in Table~\ref{tab:stellar}.

We selected the MIST stellar models \citep{choi16mist, dotter2016mist} to use for the isochrone fitting within the \texttt{isochrones} module. The derived stellar parameters from this modelling are given in Table~\ref{tab:stellar}.\@ From this analysis, we found \Nstar\ to be a slightly evolved G-type star with a mass and radius of \mstar\,=\,\Nstarmass\,\msun\ and \rstar\,=\,\Nstarradius\,\rsun.

Using the wavelet analysis, we derived a rotational velocity of \vsini\ $< 0.5$\,\kms. We also check for evidence of rotational modulation in the NGTS photometry by applying a Lomb-Scargle algorithm \citep{lomb76, scargle82} to the \NGTS\ photometric residuals.  No significant rotation modulation was detected, indicating that \Nstar\ has a low degree of stellar activity. This is consistent with the low rotational velocity (\Nvsini\,\kms) and the age of \Nstar\ derived from the \texttt{isochrones} fit (\Nage\,Gyr).

\begin{table}
	\centering
	\caption{Stellar Properties for \Nstar}
	\begin{tabular}{lcc} 
	Property	&	Value		&Source\\
	\hline
	\multicolumn{3}{l}{Identifiers}\\
	\Nstar \\
	\Nticid \\
	GAIA DR2\,\NGAIAid \\
	\Ntwomass \\
    \hline
    \multicolumn{3}{l}{Astrometric Properties}\\
    R.A.		&	\NRA			&{\em Gaia}	DR2\\
	Dec.		&	\NDec			&{\em Gaia}	DR2\\
    $\mu_{{\rm R.A.}}$ (\masy) & \NpropRA & {\em Gaia} DR2\\
	$\mu_{{\rm Dec.}}$ (\masy) & \NpropDec & {\em Gaia} DR2\\
	Parallax (mas)   &   \Nplx           &{\em Gaia} DR2\\
    \\
    \multicolumn{3}{l}{Photometric Properties}\\
	NGTS (mag)	&\NNmag		&This work\\
	TESS (mag)  &\NTmag     &TIC8\\
    V (mag)		&\NVmag 	&APASS\\
	B (mag)		&\NBmag		&APASS\\
    GAIA g (mag)&\NGAIAmag	&{\em Gaia} DR2\\
    GAIA B$_P$ (mag)  &\Nbpmag    &{\em Gaia} DR2\\
    GAIA R$_P$ (mag)  &\Nrpmag    &{\em Gaia} DR2\\
    J (mag)		&\NJmag		&2MASS	\\
   	H (mag)		&\NHmag		&2MASS	\\
	K (mag)		&\NKmag		&2MASS	\\
    W1 (mag)	&\NWmag		&WISE	\\
    W2 (mag)	&\NWWmag	&WISE	\\
    \\
    \multicolumn{3}{l}{Derived Properties}\\
    \teff\ (K)    & \Nteff               &Sec.~\ref{sub:stellar}\\
    \feh\	     & \Nmetal			    &Sec.~\ref{sub:stellar}\\
    \vsini\ (\kms)& \Nvsini			    &Sec.~\ref{sub:stellar}\\
    $\Gamma_{RV}$ (\ms) & \Ngamma		    &Sec.~\ref{sub:global}\\
    \logg               & \Nlogg			&Sec.~\ref{sub:stellar}\\
    \mstar (\msun) & \Nstarmass		        &Sec.~\ref{sub:stellar}\\
    \rstar (\rsun) & \Nstarradius	            &Sec.~\ref{sub:stellar}\\
    $\rho_*$ (\gccc) & \Nstardensity              & Sec.~\ref{sub:stellar}\\
    Age	(Gyr)			& \Nage				        	&Sec.~\ref{sub:stellar}\\
    Distance (pc)	&  \Ndist	                &Sec.~\ref{sub:stellar}\\
	\hline
    \multicolumn{3}{l}{2MASS \citep{2MASS};}\\
    \multicolumn{3}{l}{APASS \citep{APASS}; WISE \citep{WISE};}\\
    \multicolumn{3}{l}{{\em Gaia} DR2 \citep{GAIA_DR2}; TIC8 \citep{stassun2019tic8}}\\
	\end{tabular}
    \label{tab:stellar}
\end{table}

\begin{figure*}
	\includegraphics[width=\textwidth]{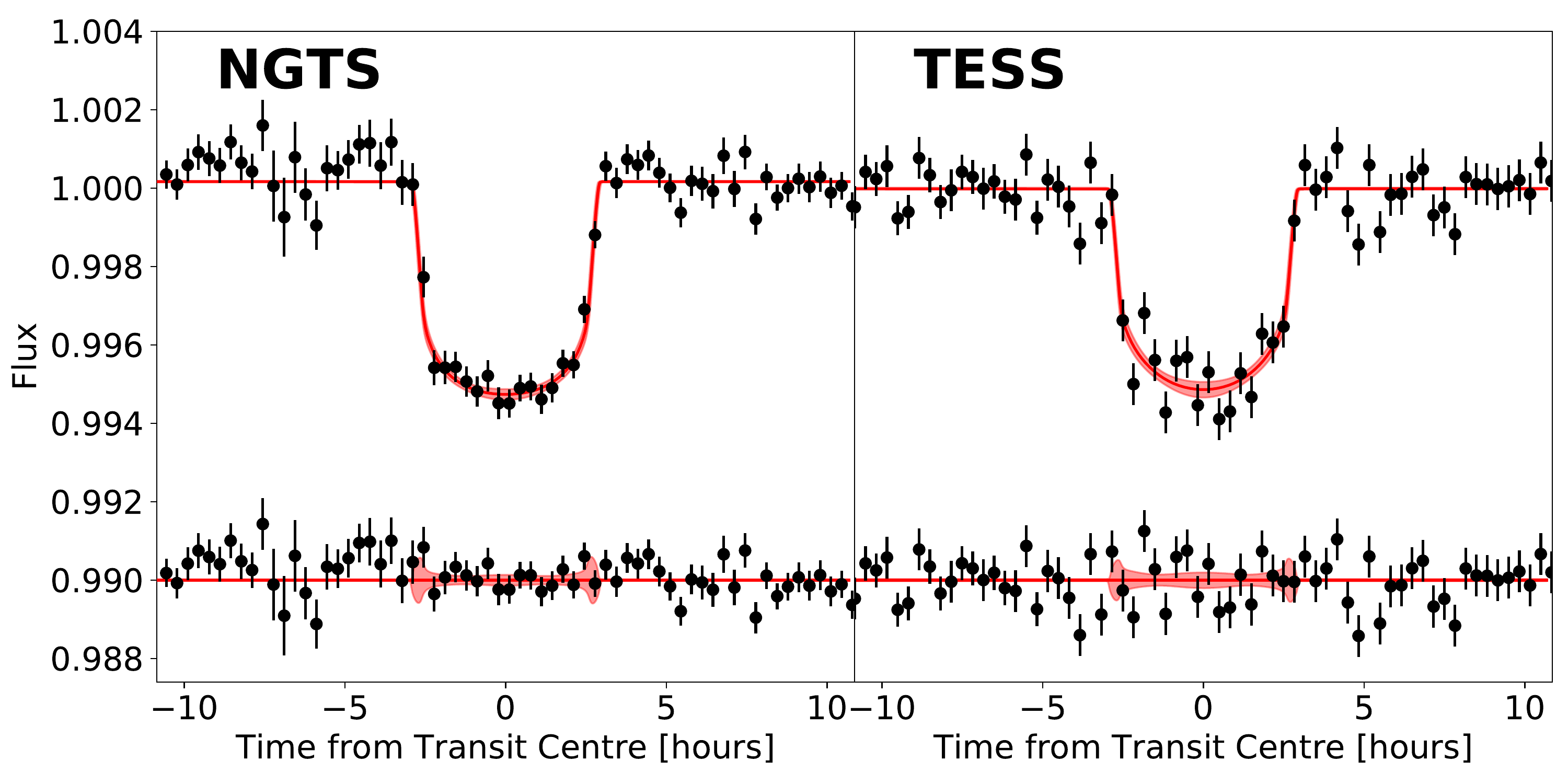}
    \caption{Phase-folded \NGTS\ and TESS light curves for \Nstar, with the models derived in Section~\ref{sub:global} plotted. The solid red lines give the median models and the shaded bands give the 1~$\sigma$ uncertainties. The residuals are plotted below the light curves in both panels. For both data sets, the flux plotted has been binned to 20\,minutes in phase.  The TESS transit depth is slightly shallower due to blending in the TESS photometry as discussed in Section~\ref{sub:dilution}.}
    \label{fig:phot}
\end{figure*}

\begin{figure}
    \centering
    \includegraphics[width=\columnwidth]{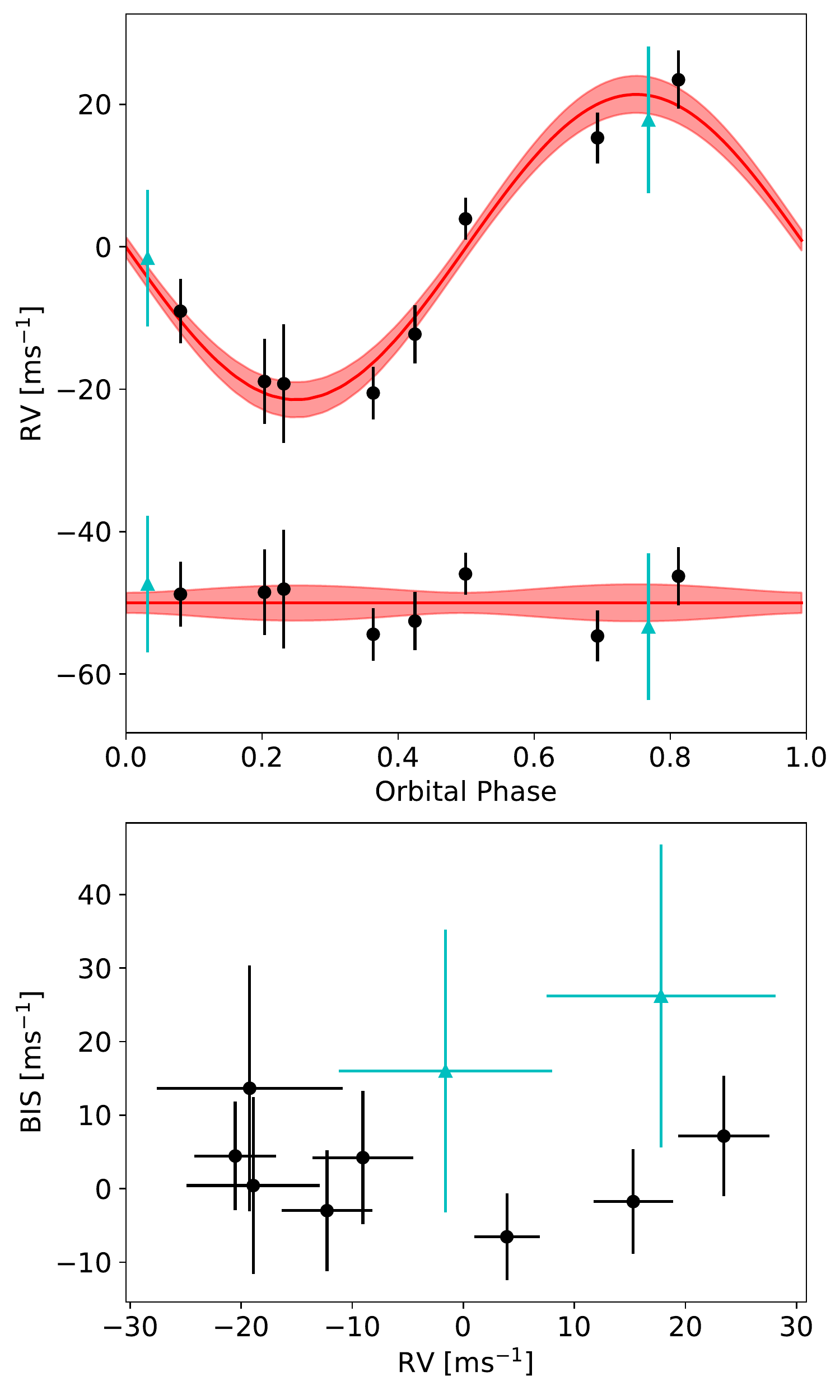}
    \caption{\textbf{Top:} Phase folded radial velocities of \Nstar\ from HARPS (black circles) and FEROS (blue triangles). The systemic $\gamma_{\rm RV}$ has been removed from the measurements and the FEROS points have had the RV offset applied. The red line gives the median RV model derived in Sec.~\ref{sub:global}, with the 1$\sigma$ tolerance on this model given by the red shading. As with Fig.~\ref{fig:phot}, the residuals are plotted offset below the RV curve. \textbf{Bottom}: Bisector spans for the HARPS and FEROS CCFs. The errorbars on the bisector spans are taken as twice the RV error for a given data point.}
    \label{fig:RV}
\end{figure}

\subsection{Global Modelling}
\label{sub:global}
We used the \textsf{exoplanet} Python package \citep{exoplanet:exoplanet} to simultaneously model the photometric and spectroscopic data. \textsf{exoplanet} is a Python toolkit which uses Hamiltonian Monte Carlo methods implemented through PyMC3 \citep{exoplanet:pymc3} to probabilistically model transit and radial velocity time series data for exoplanet systems. At each step in the sampling, we determined the Keplerian orbit of the planet around \Nstar, and used this orbit to compute the model the light curve \citep[via \texttt{starry};][]{exoplanet:luger18} and the radial velocity.

We fitted for the following planetary system parameters: time of transit centre, \tc, the orbital period, $P$, the planet-to-star radius ratio, $\rpl/\rstar$, $\ln K$ (where $K$ is the RV semi-amplitude), and the impact parameter, $b$. We constrained these parameters using the priors given in Table~\ref{tab:planet}. The stellar density, \rhostar, was also included as a free parameter, and was constrained using the value derived in Section~\ref{sub:stellar}. This was done to ensure physically realistic orbital parameters were derived. In addition, we fitted for the limb-darkening coefficients for both NGTS and TESS. We used a quadratic limb-darkening model and sampled the parameters using the efficient parameterisation of \citet{exoplanet:kipping13b}. We also fitted for the mean out-of-transit flux levels in the NGTS and TESS light curves, $F_{0,\,{\rm NGTS}}$ and $F_{0,\,{\rm TESS}}$, as well as the systemic radial velocity of \Nstar, $\Gamma_{RV}$, and the systematic instrumental radial velocity offset between HARPS and FEROS, $\Delta\,RV_{\rm FEROS}$.  From an initial modelling run, we found the formal NGTS photometric uncertainties to be slightly underestimated, so we also included an additional error scaling term, $\sigma_{\rm NGTS}$, which was added in quadrature to the formal NGTS photometry uncertainties. 

Since GAIA DR2 shows no stars within the NGTS photometric radius of \Nstar, we conclude the NGTS survey photometry is undiluted.  However from a visual inspection of the data it is evident that the TESS transits are slightly shallower depth compared with the transits in the NGTS data.  This is a result of the larger pixel scale of TESS, meaning that the TESS photometry is diluted by nearby stars, especially TIC157230670 (see Fig.~\ref{fig:tessap}).  We therefore include a dilution factor, $D$, for the TESS photometry, given by
\begin{equation}
    D \ = \ F_{cont} / F_{target},
\end{equation}
where $F_{target}$ and $F_{cont}$ are the flux contributions within the TESS photometric aperture from the target (\Nstar) and contaminating stars respectively. 

For the sampling we ran 10 chains for 4000 steps each to tune the sampler, followed by a further 7500 steps to sample the posterior distributions. The starting points for the chains were selected from the results of a least squares minimization prior to the sampling. We determined the Gelman-Rubin statistic \citep[\^R;][]{gelmanrubin92} for each chain. For all chains, we found \^R\,$\ll 1.01$, indicating that all the chains had converged, and were well-mixed. The resulting photometric and spectroscopic models are plotted in Figs.~\ref{fig:phot}~\&~\ref{fig:RV} respectively. We note that the varying levels of photometric scatter in the \NGTS\ light curve in Fig.~\ref{fig:phot} are due to the non-uniform sampling of phase space in the \NGTS\ photometry. This is because the orbital period is both long and close to a half-integer day. The median parameter values from the posterior distributions are given in Table~\ref{tab:planet}, along with the $1\sigma$ parameter uncertainties. These uncertainties are derived from the 16$^{\rm th}$ and 84$^{\rm th}$ percentile values of the posterior distributions.

We ran two models; the first with a fixed circular orbit, and the second with the orbital eccentricity, $e$, and the argument of periastron, $\omega$, allowed to freely vary. For the free eccentricity model, we constrained $e$ with the beta distribution prior parameterisation from \citet{exoplanet:kipping13, kipping2014ecc}. From this second model, we derived a value of the eccentricity of \Necc\, with a 95\% confidence upper limit of 0.08. Comparing the resultant maximum log-likelihoods (\lnL) from the two fits, and we find that allowing $e$ to freely vary results in a slightly higher \lnL. To determine whether this slight increase in \lnL\ warrants the inclusion of two additional free parameters to the model, we compare the the Bayesian Information Criterion (BIC) for the two models. The BIC is given by
\begin{equation}
    {\rm BIC} \ = \ k\,\ln\,n \ - \ 2\lnL,
\end{equation}
where $k$ is the number of free parameters in the model, and $n$ is the number of data points. A higher BIC denotes evidence against a given model. Comparing the BIC for the full global models, we find $\Delta\,$BIC = BIC$_{\rm ecc}$ - BIC$_{\rm circ}$ = 19.56. This is strong evidence in favour of the forced circular model. Comparing the BIC for just the radial velocity data for the two models, we again find a positive $\Delta\,$BIC = 4.26. In addition, we consider the significance test from \citet{lucysweeney71}, which asserts that an eccentricity measurement must satisfy $e > 2.45\,\sigma_e$, where $\sigma_e$ is the measured uncertainty in eccentricity, in order to be considered significant. Our measured value of $e = \Necc$ does not meet the criteria. Therefore, we find the orbit of \Nplanet\ to be circular at a statistically significant level, and therefore adopt the parameters derived using the fixed circular orbit for our final system parameters. 

For the large bulk of hot Jupiter planets with orbital periods $\leq$ 4\,days, we would expect a circular orbit, due to the quick circularization timescales. For example, using equation 3 from \citet{adams06circularization} with a tidal quality factor of $Q_p = 10^6$, we find a tidal circularization timescale of just 375\,Myr for a planet of Jupiter size and mass in a 4\,day orbit around a solar mass star. However, at longer orbital periods, such as the \Nperiodshort\,day orbit of \Nplanet, the situation is less clear cut; the circular orbit of \Nplanet\ is less obvious a priori. Again following \citet{adams06circularization} and with $Q_p = 10^6$, we find an orbital circularization timescale for \Nplanet\ of $\tau_{circ} =$~\Ntaucirc\,Gyr. This is significantly shorter than the lifetime of the system (\Nage\,Gyr), and so tidal circularization alone is enough to explain the circular orbit.

\subsection{Dilution of the TESS transits}\label{sub:dilution}
As discussed in Section~\ref{sub:global}, we fitted a dilution factor, $D$, for the TESS photometry, and found a value of \NdilutionMC\,\%. We expect that the majority of the contaminating flux, $F_{cont}$, in the TESS photometry will be due to the bright (T=12.016\,mag) neighbour TIC157230670. We inspected the TESS full frame image to determine the point-spread function (PSF) in the region of the camera surrounding \Nstar\ to estimate the contamination from TIC157230670, as well as other nearby stars.

We model the PSF as a 2D elliptical Gaussian using the ($S, \ D, \ K$) parameterisation PSF model \citep{pal2009}. As a part of the Cluster Difference Imaging Photometric Survey (CDIPS), \citet{bouma2019cdips} calculated ($S, \ D, \ K$) shape parameters for a selection of stars across the TESS cameras. We extract shape parameters for the sample of stars with CCD positions closest to \Nstar.  There were relatively few CDIPS targets in TESS Sector 10, Camera 2, CCD4. However, since the shape of the PSF depends almost entirely on the TESS optics, the shape parameters for a given CCD are stable across the different sectors. As such, we can use data from TESS Sector 7, for which Camera 2 CCD 4 has a significantly higher coverage of CDIPS stars, to determine the shape parameters for our model PSF for \Nstar. We therefore extract shape parameters for \Nstar\ of $S = 1.688\pm0.022$, $D = -0.0220 \pm 0.0047$, $K = 0.104 \pm 0.012$. 

We use this model PSF to estimate the amount of light from the nearby stars which falls into the photometric aperture used for \Nstar. We consider all the stars in TICv8 \citep{stassun2019tic8} within 1$\arcmin$ of \Nstar, as well as the bright neighbour TIC157230670 (see Fig.~\ref{fig:tessap}). We use the TESS mag values for each neighbouring star to weight the total flux contained within each star's PSF. This method yields an estimated dilution factor of $D =$ \NdilutionCalc\,\%, which is consistent with the dilution factor derived from the global model.  We are therefore confident that the difference in the transit depths between NGTS and TESS is due to blending in the TESS photometry, and that our global modeling is correctly accounting for this blending.

This dilution estimation method can be applied to other TESS targets where the light curve is extracted from the TESS full-frame images. Since we had undiluted photometry from NGTS for \Nstar, we simply used this as a check to ensure the fitted dilution factor (Section~\ref{sub:global}) was a physically reasonable value.  However in other cases where no undiluted photometry is available, this method can be used to estimate the degree of dilution in the TESS photometry, ensuring an accurate estimation of the planet radius.

\section{Discussion}
\label{sec:discussion}
\begin{figure}
    \centering
    \includegraphics[width=\columnwidth]{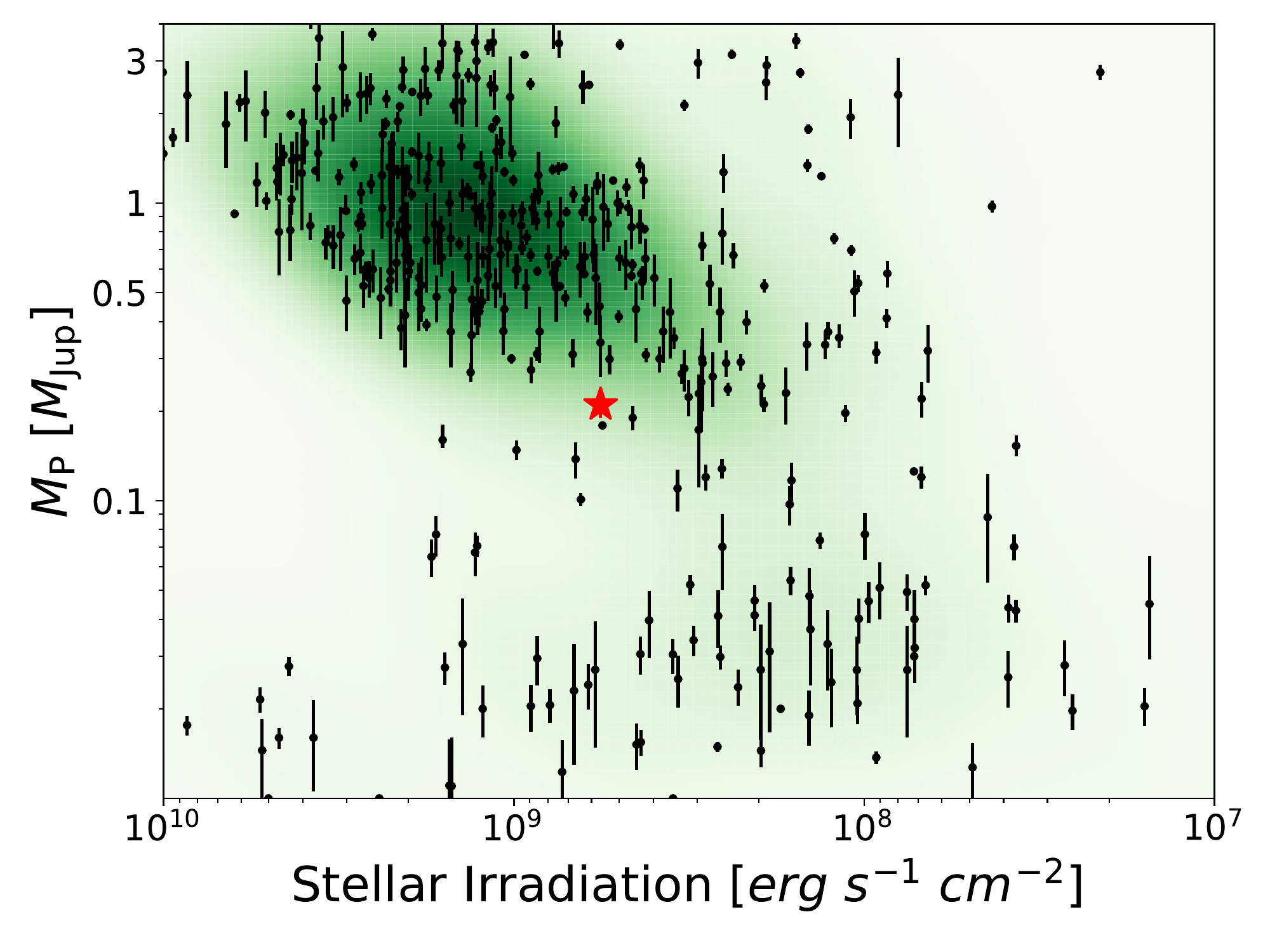}
    \caption{Irradiating stellar flux incident on the planet vs planetary mass for a selection of planets from the NASA Exoplanet Archive with a radius measured to better than 10\% and a mass measured to better than 50\%. \Nplanet\ is plotted with the red star, and the color shading gives the number density of planets in a given region of irradiation-mass parameter space.}
    \label{fig:irrvsmpl}
\end{figure}
\begin{figure}
    \centering
    \includegraphics[width=\columnwidth]{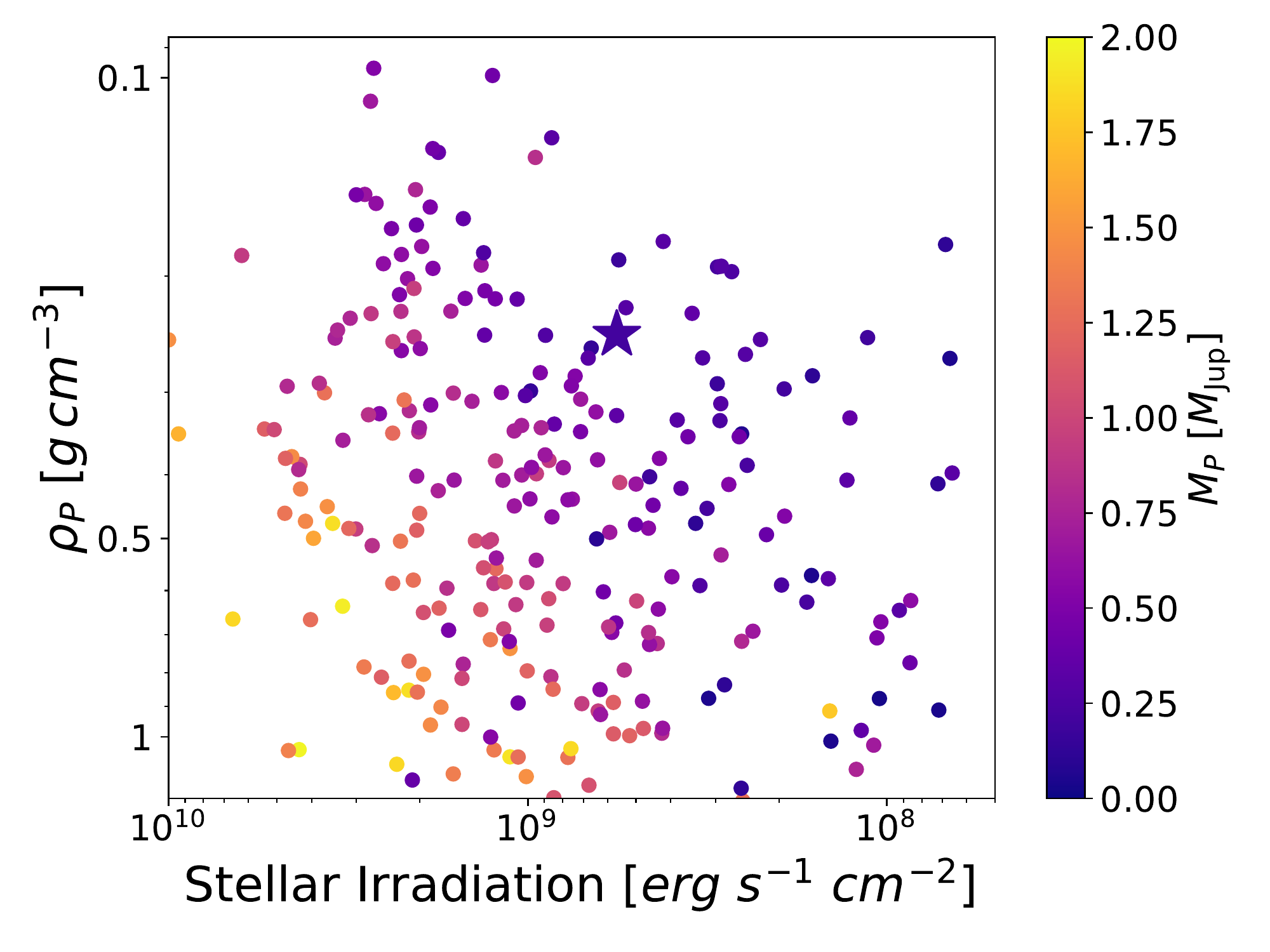}
    \caption{Distribution of the densities of known exoplanets against incident stellar flux. The circles are the same sample of planets as Fig.~\ref{fig:irrvsmpl}, but with an additional mass constraint of 0.01\,\mjup\ <= \mpl\ <= 2.0\,\mjup. The star gives the position of \Nplanet. The colour of the points gives the planetary mass. Note that both axes are inverted, so low density, highly irradiated planets reside at the top left of the plot.}
    \label{fig:irrvsrhopl}
\end{figure}
\begin{figure}
    \centering
    \includegraphics[width=\columnwidth]{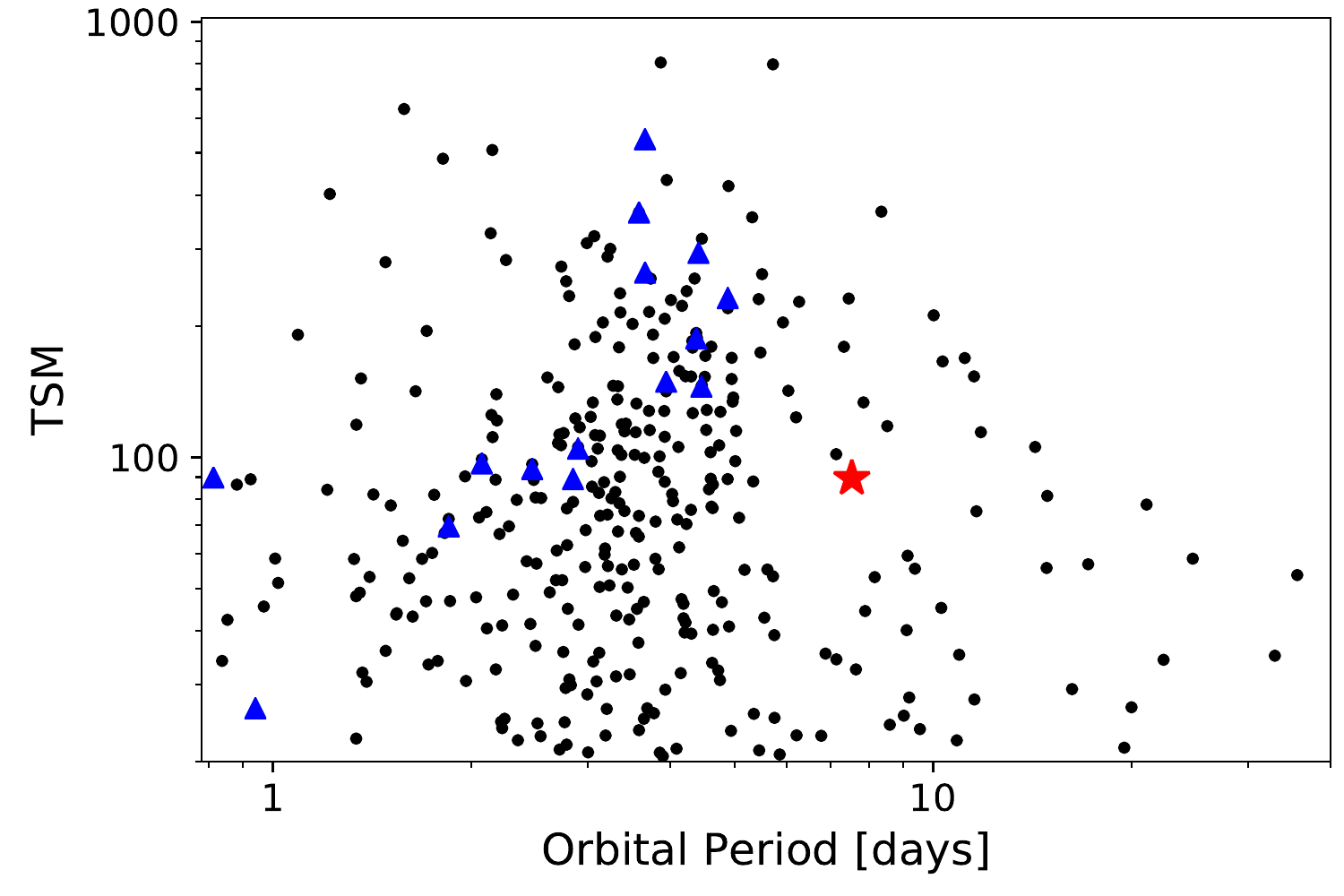}
    \caption{Transmission Spectroscopy Metric (TSM) against orbital period for the same sample of planets as Fig.~\ref{fig:irrvsmpl}. \Nplanet\ is denoted by the red star and the JWST community targets \protect\citep{stevenson2016jwstCT1, bean2018jwstCT2} are highlighted by the blue triangles.}
    \label{fig:tsm}
\end{figure}
\begin{figure}
    \centering
    \includegraphics[width=\columnwidth]{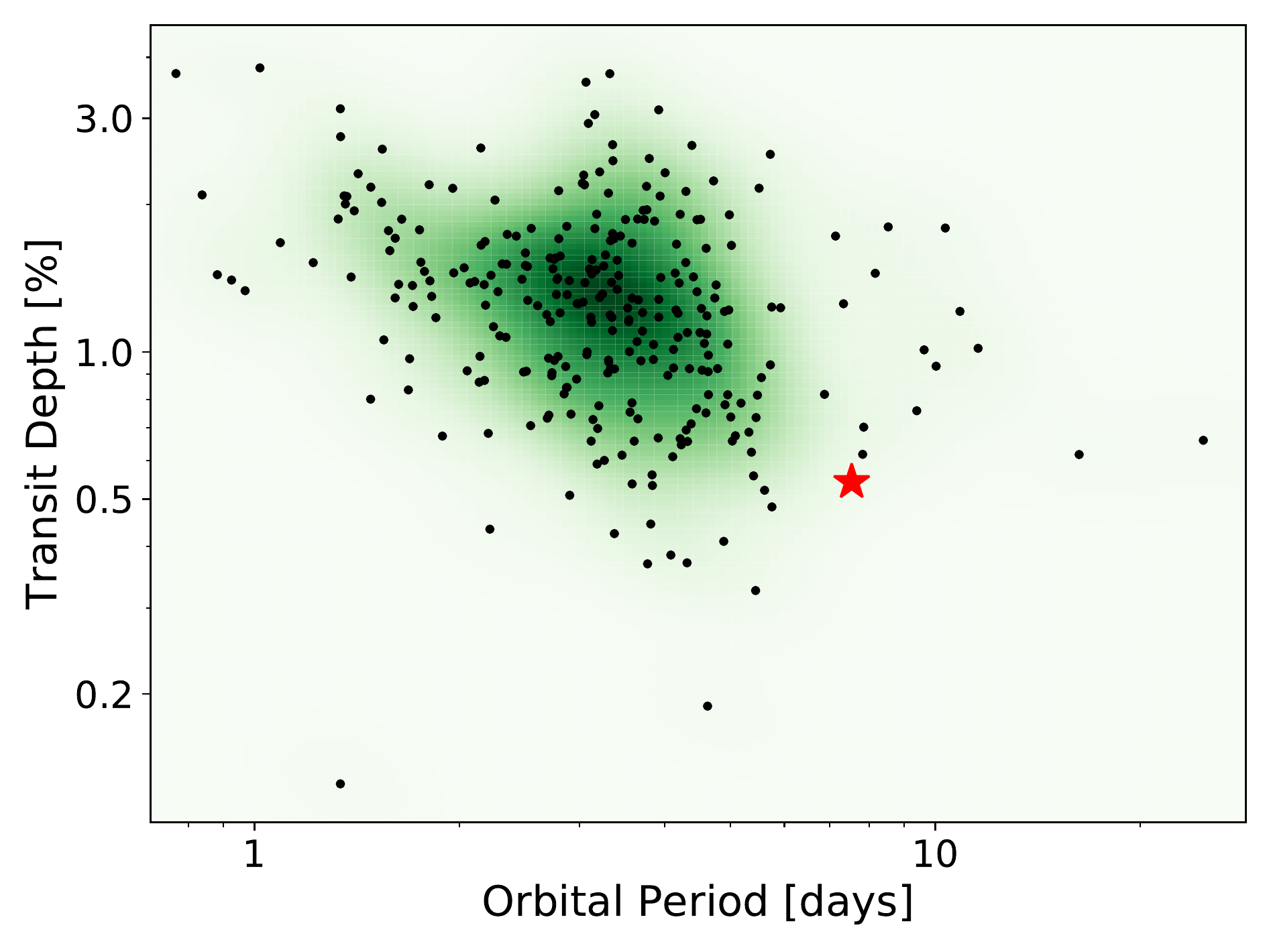}
    \caption{Orbital period and transit depth for all planets detected by ground-based transit surveys, with a radius measured to better than 10\% and a mass measured to better than 50\%. \Nplanet\ is again plotted with the red star. The colour shading gives the number density of planets in a given region of parameter space.}
    \label{fig:pervsdepth}
\end{figure}

\begin{table*}  
 \centering  
 \small  
 \caption[Model parameters]{Fitted and derived parameters (including priors) for \Nplanet\ from modelling detailed in Section~\ref{sub:global}.} 
 \label{lc_model_tab}  
 \begin{tabular}{l l l c c }  
 \noalign{\smallskip} \noalign{\smallskip} \hline  \hline \noalign{\smallskip}  
 Parameter  &   Symbol  &  Unit  &  Prior & Value \\   
 \noalign{\smallskip} \hline \noalign{\smallskip} \noalign{\smallskip}  
 \multicolumn{5}{c}{{\bf Fitted parameters}} \\   
 \noalign{\smallskip} \noalign{\smallskip} 
 Time of Transit Centre &  \tc &  BJD (TDB)  &  $\mathcal{U}\left(2458572.70, 2458572.90\right)$ & \Ntc \\
 \noalign{\smallskip} 
 Orbital Period & $P$ & days & $\mathcal{N}\left(7.5328, 0.1\right)$ & \Nperiod \\
 \noalign{\smallskip} 
 Radius Ratio    &    (\rpl\ / \rstar)    &        &    $\mathcal{U}\left(0., 1.\right)$ & \Nrratio     \\  
 \noalign{\smallskip}
 Impact Parameter & $b$   &    & $\mathcal{U}\left(0., \ \ \rpl\ / \rstar\right)$ & \Nimpact \\
 \noalign{\smallskip}
 Stellar Density & \rhostar\ &  \gccc & $\mathcal{N}\left(0.359, 0.039\right)$ & \Nstardensityglobal \\
 \noalign{\smallskip}  
 Nat. Log. of RV Semi-Amplitude & $\ln(K)$ & $\ln$ (\ms) & $\mathcal{N}\left(3.075, 3.0\right)$ & \Nlogk \\
 \noalign{\smallskip}
 Eccentricity & $e$ &  & $\beta \left(0.697, 3.27\right)^a$ & 0.0$^b$ \\
 \noalign{\smallskip}
 NGTS LDC$^c$ 1  & $U_{1,\,{\rm NGTS}}$ & & $\mathcal{K}^d\left(0., 1.\right)$  &  \NLDoneNG \\
 \noalign{\smallskip}
 NGTS LDC$^c$ 2  & $U_{2,\,{\rm NGTS}}$ & & $\mathcal{K}^d\left(0., 1.\right)$  &  \NLDtwoNG \\
 \noalign{\smallskip}
 TESS LDC$^c$ 1  & $U_{1,\,{\rm TESS}}$ & & $\mathcal{K}^d\left(0., 1.\right)$  &  \NLDoneTS \\
 \noalign{\smallskip}
 TESS LDC$^c$ 2  & $U_{2,\,{\rm TESS}}$ & & $\mathcal{K}^d\left(0., 1.\right)$  &  \NLDtwoTS \\
 \noalign{\smallskip}
 Mean NGTS Flux Baseline & $F_{0,\,{\rm NGTS}}$ &  & $\mathcal{U}\left(0.998, 1.002\right)$ & \NmeanNG \\
 \noalign{\smallskip}
 Mean TESS Flux Baseline & $F_{0,\,{\rm TESS}}$ &  & $\mathcal{U}\left(0.998, 1.002\right)$ & \NmeanTS \\
 \noalign{\smallskip}
 NGTS Error Term & $\sigma_{\rm NGTS}$ &  &  $\mathcal{N}\left(0.002, 0.1\right)$ & \NsigNG \\
 \noalign{\smallskip}
 TESS Dilution Factor & $D$ & \% &  $\mathcal{U}\left(0., 50.\right)$ & \NdilutionMC \\
 \noalign{\smallskip}
 Systemic Radial Velocity & $\gamma_{\rm RV} $  & \ms & $\mathcal{N}\left(-12595, 50\right)$ & \Ngamma \\
 \noalign{\smallskip}
 FEROS RV Offset & $\Delta\,RV_{\rm FEROS}$ & \ms & $\mathcal{N}\left(0., 50\right)$ & \Noffsetferos \\
 \noalign{\smallskip}
 \hline  
 \noalign{\smallskip} \noalign{\smallskip}
 \multicolumn{5}{c}{{\bf Derived parameters}} \\   
 \noalign{\smallskip}
 Planet Radius & \rpl\  &  \rjup\  &  &  \Nradius \\
 \noalign{\smallskip}
 Planet Mass  & \mpl\ & \mjup\ & & \Nmass \\
 \noalign{\smallskip}
 Planet Density & \rhopl\ & \gccc\ & & \Ndensitycgs \\
 \noalign{\smallskip}
 Planet Surface Gravity & $\log g_{P}$ & $\log cm\, s^{-2}$ & & \Ngravp \\
 \noalign{\smallskip}
 RV Semi-Amplitude & $K$ & $m\,s^{-1}$ & & \Nkamp \\
 \noalign{\smallskip}
 Scaled Semi-Major Axis & $a/$\rstar\ & & & \Naoverr \\
 \noalign{\smallskip}
 Semi-Major Axis & $a$ & AU & & \Nau \\
 \noalign{\smallskip}
 Orbital Inclination & $i$ & degrees & & \Ninc \\
 \noalign{\smallskip}
 Transit Duration & T$_{14}$ & hours & & \Nduration \\
 \noalign{\smallskip}
 Incident Stellar Flux & $F_{\rm inc}$ & \ergscm & & \Nflux \\
 \noalign{\smallskip}
 Equilibrium Temperature$^e$ & $T_{eq}$ & K & & \NTeq \\
 \noalign{\smallskip}
 Atmospheric Scale Height$^f$ & $H$ & km & & \NHkm \\
 \noalign{\smallskip}
 \hline
 \end{tabular}  
 \begin{list}{}{}
 \item[a  -  This is the $\beta$ distribution for the eccentricity of short period planets from \citet{exoplanet:kipping13}]
 \item[b  -  Adopted eccentricity. 95\% confidence level upper limit is 0.08]
 \item[c  -  LDC = limb darkening coefficient]  
 \item[d  -  $\mathcal{K}$ denotes the informative quadratic LDC parameterisation from \citet{exoplanet:kipping13b}]
 \item[e  -  assuming zero albedo]
 \item[f  -  assuming atmospheric mean molecular mass is 2.2u]
 \end{list} 
 \label{tab:planet}
 \end{table*} 

From the analysis set out in Section~\ref{sec:analysis}, we find \Nplanet\ to be an inflated sub-Saturn mass planet, with a mass and radius of \mpl = \Nmass\,\mjup\ and \rpl = \Nradius\,\rjup. This mass and radius gives \Nplanet\ a relatively low density of \rhopl\,=\,\Ndensitycgs\,\gccc, and using the method of \citet{southworth2007} we derive a surface gravity of $\log\,g_{P}$\,=\,\Ngravp. Based on these parameters, we expect \Nplanet\ to have an extended gaseous atmosphere.

We compare \Nplanet\ to the population of currently known planets in Fig.~\ref{fig:irrvsmpl}. We note that \Nplanet\ is at the lower mass edge of the large majority of the currently known hot-Jupiter planet population. In Fig.~\ref{fig:irrvsmpl}, the presence of the Neptunian desert \citep[eg.][]{mazeh2016} can be clearly seen, and \Nplanet\ lies on the edge of the desert. The desert boundaries found by \citet{mazeh2016} are not particularly well defined for periods $P > 5$\,days, and based on this one might not consider \Nplanet\ a true Neptune desert planet based on period alone. The fact that \Nplanet\ quite clearly lies on the boundary of the desert when we consider incident flux instead of simply period implies that the stellar parameters must be taken into account when studying the Neptune desert and the physical processes which shape it. This same effect has been investigated previously in more detail by \citet{eigmuller2019ngts5} in relation to \NGTS-5b.

Tidal interaction between the star and planet has recently been shown to be a significant source of inflation for sub-Saturn mass planets \citep{millholland2020tidalinflation}. Despite focusing on planets with \rpl $< 8.0$\rearth, \citet{millholland2020tidalinflation} also noted that tidal inflation could play a role for larger radius planets \citep[eg. WASP-107b; ][]{anderson2017wasp107}. \citet{baraffe08planetstructure} provide model radii for irradiated planets at a distance of 0.045\,AU from the Sun. This is a reasonable approximation as the irradiation received by \Nplanet\ is equivalent to orbiting at a separation of 0.049\,AU from the Sun. Assuming a heavy metal fraction of $Z = 0.02$, the \citet{baraffe08planetstructure} tables give a predicted radius for a 50\,\mearth\ planet at an age of 7.95-10.01\,Gyr of 1.044-1.057\,\rjup. This prediction agrees well with the derived radius of \Nradius\,\rjup. Therefore, we conclude that the inflation of \Nplanet\ is driven by stellar irradiation and tidal effects have very little, if any, influence on \rpl. Due to the circular orbit of \Nplanet\ this is not surprising, however obliquity tides alone could still have had an impact \citep{millholland2020tidalinflation}.

We also investigate how the bulk density of a planet, \rhopl, varies with both the stellar irradiation flux incident on it, $F_{\rm inc}$, as well as the planet's mass, \mpl. From Fig.~\ref{fig:irrvsrhopl} we can see a steady decrease in \rhopl\ with increasing $F_{\rm inc}$ for planets of a similar mass. We also note from Fig.~\ref{fig:irrvsrhopl} that \Nplanet\ has one of the lowest densities for planets with a similar level of irradiation. It can also be seen that the majority of planets with lower densities than \Nplanet\ have both a larger $F_{\rm inc}$ and \mpl. We also note that all the planets in the top left region of Fig.~\ref{fig:irrvsrhopl} with $F_{\rm inc} > 10^9$\,\ergscm\ and \rhopl\ < 0.2\,\rhojup\ have a greater mass than \Nplanet. This implies that a minimum mass is required to maintain an extended atmosphere under these extreme levels of irradiation. This suggests that the level of irradiation received by \Nplanet\ is close to an upper limit of irradiation before significant atmospheric loss occurs.

Given \Nplanet\ has a low bulk density (\Ndensitycgs\,\gccc), a relatively high equilibrium temperature (\NTeq\,K) and orbits a fairly bright host star (T=\NTmagshort), \Nplanet\ provides a valuable target for atmospheric characterization through transmission spectroscopy. \citet{kempton2018tsm} introduced a transmission spectroscopy metric (TSM) for the purposes of identifying good targets for transmission spectroscopic follow-up, based on the expected signal-to-noise. In Fig.~\ref{fig:tsm} we compare the TSM value for \Nplanet\ to other transiting planets. We note that \Nplanet\ has a reasonably high TSM, especially when compared to planets of similar and longer orbital periods. Therefore, \Nplanet\ presents an opportunity to study the atmosphere of a planet on a longer period than the main population of currently known hot Jupiters.

The discovery of \Nplanet\ again shows that NGTS is able to find transiting planets that were not detectable to previous ground-based transit surveys. Fig.~\ref{fig:pervsdepth} shows that \Nplanet\ is one of the longest period planets discovered with a ground-based transit survey to date. Only 12 other ground-based transit survey planets have longer orbital periods. Of these, all have deeper transits than \Nplanet. The placement of \Nplanet\ on in Fig.~\ref{fig:pervsdepth} suggests that the higher photometric precision achieved by \NGTS\ compared to other ground-based transit surveys will allow more planets with longer orbital periods and shallow transits to be found by \NGTS\ over the coming years.

\subsection{Fate of \Nplanet}\label{sub:fate}

The age of \Nstar\ (\Nage\,Gyr), along with its Solar-like mass (\Nstarmass\,\msun) and metallicity ([Fe/H] = \Nmetal), indicates that it will soon, or already has, left the main sequence. Indeed, within these uncertainties, the star may either still reside on the main sequence or is already slightly evolved. The increased stellar radius of \Nstarradius\,\rsun\ for a mass of \Nstarmass\,\msun\ is evidence for the latter. Therefore, we expect that the expansion of \Nstar\ will begin to increase the irradiation flux received by \Nplanet. 

By analogy with the fate of the Sun \citep{schsmi2008,veras2016a,veras2016b}, upon leaving the main sequence, \Nstar\ will enlarge and tidally draw in and destroy planets which are separated by a distance less than about 1\,AU. For \Nplanet, with a semi-major axis under 0.08\,AU, the question is not if the planet will be engulfed, but when. Here, we estimate when and where the engulfment will take place.

In post-main-sequence planetary systems, the inward tidal force \citep{musvil2012,adablo2013,madetal2016} competes against the outward orbital expansion due to mass loss from stellar winds \citep{hadjidemetriou1963,veretal2011}. However, with such a small semi-major axis (\Nau\,AU), \Nplanet\ will be subjected to negligible orbital expansion before tidal forces take hold. Further, our current knowledge of the system indicates that no other forces need to be considered in the computation. Even evaporation of the planet's atmosphere during the latter stages of \Nplanet's lifetime might be negligible enough to not substantially affect its orbital evolution \citep{schetal2019}.

Given these assumptions, we use the tidal equations from \cite{viletal2014} to compute the engulfment time; these tides are often termed "dynamical tides", originally from \cite{zahn1977}. A key element in the computation is the stellar model adopted, which provides a time series in the mass and radius values in the different components in the host star. Given the extent of the uncertainties in stellar age and metallicity, we are content with just a rough estimate. Hence, we use the {\tt SSE} stellar evolution code \citep{huretal2000} with an initial stellar mass of 1.021\,\msun\ and an initial value of $Z = 0.0133$ (computed from $Z = 10^{[{\rm Fe/H}]} Z_{\odot}$, with $Z_{\odot} = 0.0142$). By the time that the star evolves to an age of 9.4\,Gyr, it will have lost just $3.3 \times 10^{-4}$\,\msun.

We find that the engulfment occurs at a stellar age of 9.9358\,Gyr. However, a more sensible way of expressing this estimate is roughly 500\,Myr after the current time. The engulfment occurs when the star's radius expands out to about 0.02\,AU, which is nearly one quarter of the planet's current separation. Due to the fairly large uncertainty on the age of \Nstar\ of $\pm\,1.4$\,Gyr, we should be careful not draw too strong conclusions from these calculations. However, it is clear that the environment in which \Nplanet\ finds itself will be undergoing quite extreme changes in the not too distant future.

\section{Conclusions}
\label{sec:conc}
We report the discovery of \Nplanet, a sub-Saturn mass transiting exoplanet with a mass and radius of \Nmass\,\mjup and \Nradius\,\rjup. The orbital period of \Nplanet\ (\Nperiodshort\,days), which is longer than the periods of 95\% of the other planets discovered by all ground-based transit surveys, makes it the longest period planet to date discovered by the main \NGTS\ survey. The transits of \Nplanet\ are also detected in the full frame TESS images, and the overall precision of the phase-folded TESS light curve is similar to that of the NGTS light curve (see Fig.~\ref{fig:phot}).

With its low density (\Ndensitycgs\,\gccc), and long orbital period (\Nperiodshort\,days), \Nplanet\ represents a good chance to study the atmosphere of a planet with a longer period than the bulk population of hot Jupiters. 

The mass and irradiation of \Nplanet\ places it on the boundary of the Neptunian desert in a currently under-sampled region of parameter space. Therefore, \Nplanet\ can play an important role in understanding the physical processes shaping this region. 

\section*{Acknowledgements}
Based on data collected under the NGTS project at the ESO La Silla Paranal Observatory.  The NGTS facility is operated by the consortium institutes with support from the UK Science and Technology Facilities Council (STFC)  project ST/M001962/1. 

This study is based on observations collected at the European Southern Observatory under ESO programmes 0104.C-0588.\\ 

This research made use of \textsf{exoplanet} \citep{exoplanet:exoplanet} and its dependencies \citep{exoplanet:agol19, exoplanet:astropy13, exoplanet:astropy18, exoplanet:kipping13, exoplanet:kipping13b, exoplanet:luger18, exoplanet:pymc3, exoplanet:theano}.

We  thank  the  Swiss  National  Science  Foundation  (SNSF) and the Geneva University for their continuous support to our planet search programs. This work has been in particular carried out in the frame of the National Centre for Competence in Research {\it PlanetS} supported by the Swiss National Science Foundation (SNSF). \\ 
This publication makes use of The Data \& Analysis Center for Exoplanets (DACE), which is a facility based at the University of Geneva (CH) dedicated to extrasolar planets data visualisation, exchange and analysis. DACE is a platform of the Swiss National Centre of Competence in Research (NCCR) PlanetS, federating the Swiss expertise in Exoplanet research. The DACE platform is available at \url{https://dace.unige.ch}. \\ 

Contributions at the University of Geneva by DB, FB, BC, LM, and SU were carried out within the framework of the National Centre for Competence in Research "PlanetS" supported by the Swiss National Science Foundation (SNSF). 

The contributions at the University of Warwick by PJW, RGW, DLP, FF, DA, BTG and TL have been supported by STFC through consolidated grants ST/L000733/1 and ST/P000495/1. 

DV and DJA gratefully acknowledge the support of the STFC via Ernest Rutherford Fellowships (grant ST/P003850/1 and ST/R00384X/1).


CAW acknowledges support from the STFC grant ST/P000312/1.
TL was also supported by STFC studentship 1226157.
JIV acknowledges support of CONICYT-PFCHA/Doctorado Nacional-21191829.
JSJ acknowledges support by Fondecyt grant 1161218 and partial support by CATA-Basal (PB06, CONICYT).
MNG acknowledges support from MIT's Kavli Institute as a Juan Carlos Torres Fellow.
EG gratefully acknowledges support from the David and Claudia Harding Foundation in the form of a Winton Exoplanet Fellowship.
This project has received funding from the European Research Council (ERC) under the European Union's Horizon 2020 research and innovation programme (grant agreement No 681601).
The research leading to these results has received funding from the European Research Council under the European Union's Seventh Framework Programme (FP/2007-2013) / ERC Grant Agreement n. 320964 (WDTracer).

\section*{Data Availability}
The data underlying this article are available in the article and in its online supplementary material.




\bibliographystyle{mnras}
\bibliography{paper} 








\bsp	
\label{lastpage}
\end{document}